\documentclass[
    ,final            
  ]
  {aipproc}

\layoutstyle{6x9}

\newcommand{\be}{\begin{equation}}
\newcommand{\ee}{\end{equation}}
\newcommand{\bea}{\begin{eqnarray}}
\newcommand{\eea}{\end{eqnarray}}
\newcommand{\ba}{\begin{array}}
\newcommand{\ea}{\end{array}}

\def\beaa{\begin{array}}
\def\eeaa{\end{array}}

\begin{document}

\title{Non-decoupling effects of SUSY in the physics 
 of Higgs bosons and their phenomenological implications}

\author{Ana M. Curiel, Mar\'{\i}a J. Herrero\footnote{Talk given at the X Mexican
School of Particles and Fields, Playa del Carmen, M\'exico, 2002.} and David Temes}{
address={Universidad Aut\'onoma de Madrid}}

\begin{abstract}
We consider a plausible scenario in the Minimal Supersymmetric Standard Model (MSSM) 
where all the genuine supersymmetric (SUSY) particles are heavier than the electroweak
scale. In this situation, indirect searches via their radiative corrections to
low energy observables are complementary to direct searches, and they can be 
crucial if the SUSY masses are at the TeV energy range. We summarize the most
relevant heavy SUSY radiative effects in Higgs boson physics and emphasize those
that manifest a non-decoupling behaviour. We focus, in particular, on the 
SUSY-QCD non-decoupling effects in fermionic Higgs decays, flavour changing
Higgs decays and Yukawa couplings. Some of their phenomenological implications 
at future colliders are also studied.     
\end{abstract}

\maketitle


\section{Introduction}

The search for SUSY particles via their indirect  signals at
colliders 
can be of great relevance in the future. The indirect SUSY searches via their 
effects on radiative corrections to low energy observables are 
complementary to direct searches, and can be crucial if the SUSY particles  
turn out to be too heavy as to be produced directly. Remember that it was the
case in the past regarding the top quark indirect searches at LEP, via radiative corrections, 
which were an important guidance towards its final dicovery at the Tevatron 
collider. 

We consider here a plausible scenario in the MSSM where all the genuine SUSY particles are heavier than the electroweak
scale $m_{EW}$. In this case, the most promising indirect SUSY signals come from SUSY non-decoupling effects in low energy observables.
We shortly review here some of these most relevant effects and show that
they will provide sizeable signals at future colliders even if the SUSY masses
are as heavy as $M_{SUSY}\sim {\cal O}(TeV)$. We study the consequences for 
Higgs boson physics where the SUSY radiative corrections are known to be
sizeable, and focus on the SUSY-QCD radiative
corrections which are the dominant SUSY contributions. 

The
{\it heavy SUSY scenario} is defined by taking the soft SUSY breaking mass
parameters of the MSSM very large as compared to the electroweak scale. For
shortness, we 
choose here the simplest case where all these mass parameters are
quasi-degenerate, 
and we refer to them by a genereric common SUSY scale named $M_{SUSY}$.
Concretely, the heavy SUSY scenario is implemented by,
$M_{SUSY} \sim M_{\tilde Q} \sim M_{\tilde U} \sim M_{\tilde D} 
\sim M_{\tilde g} \sim |\mu| \gg m_{EW}$,
where $M_{\tilde Q}$,  $M_{\tilde U}$ and  $M_{\tilde D}$ characterize the heavy
squark masses, and  $M_{\tilde g}$ the heavy gluino masses. The large $|\mu|$ values,
although not needed to get heavy particles in the SUSY-QCD sector, 
guarantee (together with large values of the other relevant parameters in the 
SUSY-Electroweak sector, $M_{1,2 }$) that the charginos and neutralinos are 
also heavy.
  
In addition, we have also considered the possibility that the extra (non-SM
like) Higgs
particles get also very heavy. This is called here the {\it heavy Higgs sector
scenario} and is defined by taking the speudoscalar mass very large, 
$m_A \gg m_{Z}$. The extra Higgs particles of the
MSSM, $H^o$, $H^{\pm}$, and $A^o$ get very heavy,  
$m_{H^o} \sim m_{H^{\pm}} \sim m_A \gg m_{EW}$,
while $h^0$ remains light with a mass close to its upper bound,
$m_{h^o} \leq 135 \, \, GeV$, and with tree level interactions being 
as the Standard Model (SM) Higgs boson ones. 

The outline of this talk is as follows. First we review the concepts of  
 decoupling/non-decoupling of heavy particles. Next we review the  
 non-decoupling effects of squark-gluino loops in flavour preserving Higgs 
 decays and in flavour changing Higgs decays. Then, we summarize 
 the results for the corresponding induced Yukawa effective couplings 
 after the integration, at the one-loop level, of
 the squarks and gluinos  in the path integral. Finally, we shortly review
 the main phenomenological consequences for future colliders.

\section{Decoupling/Non-decoupling of heavy particles} 

Commonly, there are two different languages used to define the
decoupling/non-decoupling behaviour of heavy particles. One uses the concept
of the effective action (or, equivalently, the one-particle irreducible Green functions) that is generated after integration in the 
path integral of the heavy particles, and the other one studies the effects 
on the low energy 
observables that are induced from the heavy particles. To explain this more
clearly  we include next some illustrative examples.   

{\bf (a) Effective action/One-particle irreducible functions:}

Let us consider first the simple example of QED and study the effects of heavy
electrons on the physics of photons at low energies.
One starts with the classical action for photons and electrons in QED, 
$S_{QED}= -\frac{1}{4} \int dx  {F_{\mu \nu}F^{\mu \nu}} + 
\int dx \, \bar \psi_e (i \slash {\hspace{-0.3cm}D} -  {M_e})\psi_e $,
and study the effective action that is got after integration of the electron
field at one-loop level, $\Gamma_{eff} = \frac{-1}{4} \int dx  {F_{\mu \nu} F^{\mu \nu}} - \frac{e^2}{3 (4\pi)^2} \hat \Delta
\int dx  {F_{\mu \nu}F^{\mu \nu}} - \frac{e^2}{15 (4\pi)^2  {M_e^2}} \int F_{\mu \nu} \partial^2 F^{\mu \nu} + \dots$
where $\hat \Delta \equiv \frac{2}{\epsilon} + log4\pi - \gamma -
log\frac{M_e^2}{\mu^2_0}$ contains the divergent piece in four dimensions,
and an expansion in inverse powers of the heavy electron mass, $M_e \gg p$, has
been performed. The remaining terms in this expansion are suppressed by higher
inverse powers of $M_e$ and are not shown.  Notice that the second term has the same 
structure as the kinetic term of photons and, therefore its effects are not
physically observable since they can be absorbed by a photon wave function 
redefinition.  The other terms, being proportional to inverse powers of $M_e $,
vanish in the asymptotically large electron mass limit and, therefore, their effects 
decouple at low energies. This is referred to as decoupling of electrons {\it a
la} Appelquist-Carazzone, since it follows the general behaviour described in
the Decoupling Theorem by these authors~\cite{Carazzone}.

The second example refers to the decoupling of SUSY particles, for the
previously introduced heavy SUSY scenario,  
in the low energy self-interactions  of the electroweak gauge bosons,
$\gamma,Z,W$, at the one-loop level.
By using this effective action language, it has been shown in
~\cite{Dobado_MH1} that after the integration of all the SUSY particles 
(squarks, sleptons, charginos and neutralinos) in the path integral 
formalism, and by considering the large $M_{SUSY}$ limit, their effects  
are absorbed in redefinitions of the gauge boson wave functions and 
weak boson masses or else they are suppressed by inverse powers of $M_{SUSY}$.
Therefore, the SUSY particles decouple. Similarly, it has been 
shown~\cite{Dobado_MH1}  that the
effects of the extra Higgs bosons, in the previously introduced heavy Higgs
sector scenario, also decouple in the $\gamma,Z,W$ self-interactions at 
one-loop level. The low energy interactions left are then just those of the SM.

The third example is the decoupling of the extra Higgs particles in this same 
heavy Higgs sector scenario, in the self-interactions of the
lightest MSSM Higgs particle, $h_0$, at the one-loop level. It has been shown
in ~\cite{0208014} that all the effects of the heavy Higgs bosons are absorbed
by a mass correction, $\Delta m_{h_o}$, or else they are suppressed by inverse
powers of the heavy
mass $m_A$. Similarly,  it has been shown~\cite{Hollik_penar1} that the
effects of heavy stops also decouple in the $h_0$
self-interactions, at the one-loop level. These low energy $h_0$ self-
interactions, therefore, have the
same structure as the SM Higgs boson self-interactions, even at the one-loop level.

{\bf (b) Observables:} 

The first example is the non-decoupling of top quark loops in the partial decay width
$\Gamma (Z \to \bar bb)$, within the SM context. The result to one-loop level and in the 
large $m_t$ limit can be written as,
$\Gamma (Z \to \bar bb) = \Gamma_{0} \left(1 + a \frac{\alpha}{4\pi}\frac{
{m_t^2}}{m_W^2} \right)$,
where we have factorized out the treel level result and $a$ is a numerical factor.  
We see clearly that the
correction grows quadraticaly with the top mass, indicating a non-decoupling
behaviour. 

The second example is the decoupling of squark-gluino loops in the partial decay
width $\Gamma (t \to W^+ b)$. The result to one-loop level and in the large 
$M_{SUSY}$ limit can be written as,
$\Gamma (t \to W^+ b) = \Gamma_{0} 
\left(1 + b \frac{\alpha_S}{4\pi}\frac{m_t^2}{ {M_{SUSY}^2}} \right)$,
where we have factorized again the treel level result and $b$ is a numerical
factor. For asymptotically large  $M_{SUSY}$ values the correction vanishes,
indicating a decoupling behaviour.

In the following we will present some other interesting examples, within 
the MSSM, 
where the SUSY particles of the SUSY-QCD sector do not decouple. These are the 
fermionic decay widths of the Higgs bosons, both flavour preserving and flavour
violating, and the effective low energy Yukawa interactions.      

\section{Non-decoupling of SUSY loops in Higgs decays}
 
We choose the Higgs decays into quarks, $H \to q \bar q$, because, 
on one hand, these are
the dominant decays in a large region of the MSSM parameter space and, on
the other hand, 
the associated Yukawa couplings, $\lambda_{Hqq}$, enter in relevant Higgs 
production processes. We study in particular $h^o \to  \bar bb$ and 
$H^+ \to t\bar b$, because $h^0$ will be problably the first to be detected, 
and
because the charged Higgs provides itself an unambiguous signal of physics 
beyond SM.
As we have said, we focus here on the SUSY radiative corrections from
loops of squarks and gluinos. These being $\mathcal{O}(\alpha_S)$ are the 
dominant SUSY corrections.

There are two types of one-loop diagrams contributing to $h^o \to b \bar b$ 
, the triangular vertex diagrams with two sbottoms and one gluino in the internal propagators of the triangle, and the self-energy diagrams for the external bottom legs with one sbottom and one gluino in the internal propagators. These SUSY-QCD (SQCD) corrections and the genuine QCD 
corrections modify the tree level result $\Gamma_0$ by, 
$\Gamma (h^o \to b \bar b) \equiv \Gamma_0(h^o \to b \bar b)
        (1 + 2  {\Delta_{QCD}} + 2  {\Delta_{SQCD}})$,
and are known to be quite sizeable, $2\Delta_{QCD} \simeq -50\%$,
$|2\Delta_{SQCD}| \leq 50\%$.
The result of $\Delta_{SQCD}$ 
is~\cite{Dabelstein95},\\
${\Delta_{SQCD}} = \frac{\alpha_S}{3\pi}  {g_{h^o \tilde b_a \tilde b_b}} 
\frac{m_W \cos\beta}{2 g m_b \sin\alpha} 
\left[ m_b (R_{2b}^{(b)} R_{2a}^{(b)*}+ R_{1b}^{(b)} R_{1a}^{(b)*}) 
 {C_{11}} + 
 {M_{\tilde g}} (R_{2b}^{(b)} R_{1a}^{(b)*} +\right.$\\
 $ \left. +R_{1b}^{(b)} R_{2a}^{(b)*})  {C_0} \right] (m_b^2,m_{h^o}^2,m_b^2,M_{\tilde g}^2,M_{\tilde b_b}^2,M_{\tilde b_a}^2) 
+ {\Sigma}_{S}^b (m_b^2) - 2 m_b^2 \left[ {\Sigma}_{S}^{b'} (m_b^2) + 
{\Sigma}_{V}^{b'}(m_b^2) \right] $

Here $R^{(b)}$ are the rotation matrices that realte the interaction eigenstates and the sbottom mass eigenstates. Notice that the trilinear soft breaking parameter $A_b$ and $\mu $ enter in the 
$g_{h^o \tilde b_a \tilde b_b}$ couplings (not shown explicitly here, 
for brevity) and in 
 $L-R$ squark mixing. $M_{\tilde g}$ 
enters both in external factors and in the one-loop 
integrals for the vertex corrections, $C_0, C_{11}$, and for the self-energies 
$B_0 \, , \, B_1$. $M_{\tilde q}$ enters in the integrals.
The relevant $\tan \beta$ parameter enters in 
 $L-R$ squark mixing. 
 Similar results have been obtained for $H^+ \to t \bar b$ from  loops of stops, sbottoms, and gluinos~\cite{Jimenez96}.

In order to study the consequences of the heavy SUSY scenario on the previous 
Higgs boson decays, we perform expansions of the integrals and mixing angles 
appearing in $\Delta_{SQCD}$,
which are valid for $M_{SUSY}\gg m_{EW}$, and get the two first terms being 
${\cal O}(\frac{m_{EW}^2}{M_{SUSY}^2})^n$ with  $n=0,1$ respectively~\cite{Haber_Temes}. 
We give here the result for the
leading term,  $n=0$, that is valid for all $\tan\beta$ values,\\
${\Delta_{SQCD}}
        = \frac{\alpha_s}{3\pi} 
        \left\{ \frac{- {\mu M_{\tilde g}}}{ {\tilde M_S^2}}
        \left(  {\tan\beta} + \cot\alpha \right)
        f_1(R)  + {\mathcal O} \left(\frac{m_{EW}^2}{ {M_{SUSY}^2}}\right) 
        \right\}$
where, $\tilde M_S^2 \equiv \frac{1}{2}(M_{\tilde b_1}^2 + M_{\tilde b_2}^2)$,
 $R \equiv M_{\tilde g}/\tilde M_S$, $f_1(1)=1$.
We see clearly that for large $M_{\tilde g}\sim {\tilde M_S}\sim \mu \sim M_{SUSY}$, the correction
gives a non-vanishing constant and therefore the heavy squarks and gluinos
do not decouple. We also see that this non-decoupling contribution is enhanced 
at large $\tan\beta$, which is in agreement with the result in the zero
external momentum approximation and large $\tan\beta$ limit of ~\cite{Carena}.
Finally, if we also consider the heavy Higgs sector scenario, 
 $ {m_A} \gg m_Z$, where $ \cot\alpha = - {\tan\beta} - 2 \, m_Z^2 \, {\tan\beta}\cos 2\beta \, / m_A^2 + \dots $, we see that the leading term cancels and the correction decouples as $\sim m_Z^2 / m_A^2$, recovering the SM result, as expected.

%

For the case of 
$H^+ \to t \bar b$, we have found similar 
results~\cite{Penaranda_Temes}.
In particular the result for the leading term,  $n=0$, valid for all 
$\tan\beta$ values, reads,\\
${\Delta_{SQCD}}
        = \frac{\alpha_s}{3\pi} 
        \left\{ \frac{- {\mu M_{\tilde g}}}{ {\tilde M_S^2}}
        \left(  {\tan\beta} + \cot\beta \right)
        f_1(R)  + {\mathcal O} \left(\frac{m_{EW}^2}{ {M_{SUSY}^2}}\right) 
        \right\}$
where, $\tilde M_S^2 \equiv \frac{1}{2}(M_{\tilde q_1}^2 + M_{\tilde q_2}^2)$, 
 ${\tilde q} ={\tilde t},{\tilde b}$, 
 $R \equiv M_{\tilde g}/\tilde M_S$, $f_1(1)=1$,
and it agrees again with the result of the effective approach,
~\cite{Carena2}, at large $\tan\beta$.  
%
We find again the same non-decoupling behaviour as in
 $h^0 \rightarrow b \bar b$  but the correction is numerically larger. For instance,
 $\Delta_{SQCD} \simeq -40\%$ for 
 $\tan\beta=30 \, , \, M_{SUSY} = 1$ TeV. In addition, we find that 
 the next to leading order terms, with $n=1$, are smaller  than $1\%$ for
 $M_{SUSY}\ge 300\,GeV$.  

\section{Flavour changing neutral Higgs decays from squark-gluino loops}
 Flavour Changing Neutral Current (FCNC) processes are ideal to look for indirect 
SUSY signals, or any other radiative effects from possible physics beyond the SM, since the SM predicts negligible rates for these processes. These FC interactions are absent at the tree-level in the SM and in the
MSSM, but they can be generated at the one-loop level and lead to sizable contributions in the MSSM, specially at large $\tan \beta$.

 We study here the Flavour Changing Neutral Higgs Boson Decays (FCHD) generated from squark-gluino loops in the MSSM~\cite{Curiel2}. They are proportional to $\mathcal{O}(\alpha_S)$ and therefore dominate the SUSY corrections. We focus on the particular decays,
 $h_0,H_0,A_0 \rightarrow b\bar s, s \bar b$ and
 $H_0,A_0 \rightarrow t\bar c, c \bar t$
in the non-minimal flavour scenario where there is squark mixing from misalignment between quark and squark mass matrices, which constitutes the most general case in the MSSM. Our study is devoted to the second and third generation quarks because the squark mixing between these two generations is the less constrained experimentally.

 Once the quark mass matrices are diagonalized, and assuming that FC squark mixing is significant only in LL entries, the squark mass matrices for the up and down sector can be written as follows,  

\hspace{-1cm}
$ M^2_{\tilde u} =\left\lgroup  
         \beaa{llll} 
          M_{L,c}^2  &  m_c X_c &  {\Delta_{LL}^u} & 0 \\
          m_c X_c   &  M_{R,c}^2  &  0 &  0 \\
           {\Delta_{LL}^u}&  0 & M_{L,t}^2  &  m_t X_t \\
           0 &  0 & m_t X_t &  M_{R,t}^2 
\eeaa 
        \right\rgroup$, 


\vspace{-2.3cm}
\hspace{7cm}
$M^2_{\tilde d} =\left\lgroup 
         \beaa{llll}
          M_{L,s}^2  &  m_s X_s &  {\Delta_{LL}^d}&  0 \nonumber \\ 
          m_s X_s   &  M_{R,s}^2  &  0   &   0 \nonumber \\
           {\Delta_{LL}^d} &  0 & M_{L,b}^2  &  m_b X_b \nonumber\\
           0 & 0 & m_b X_b &  M_{R,b}^2 \nonumber
\eeaa
         \right\rgroup$
\begin{center}
where,
\end{center}
$M_{L,q}^2 =  {M_{\tilde Q,q}^2} +m_q^2 + \cos2\beta (T_3^{q}-Q_q s_W^2)m_Z^2 \, ,
M_{R,(c,t)}^2 =  {M_{\tilde U,(c,t)}^2} +m_{c,t}^2 + \cos2\beta Q_t s_W^2 m_Z^2\, ,$\\ 
$M_{R,(s,b)}^2 =  {M_{\tilde D,(s,b)}^2} +m_{s,b}^2 + \cos2\beta Q_b s_W^2 m_Z^2 \, , 
X_{c,t} = m_{c,t} (  {A_{c,t}} -  {\mu} \cot \beta) \, ,$\\ 
$X_{s,b} = m_{s,b} (  {A_{s,b}} -  {\mu} \tan \beta) \, ,  \,{\Delta_{LL}^u} = {\lambda} M_{L,c} M_{L,t} \, , \, 
 {\Delta_{LL}^d} =  {\lambda} M_{L,s} M_{L,b} \, ;\, { 0 \le \lambda \le 1 }$\\



 The squark mass eigenstates, $\tilde q_{\alpha}$, and the interaction eigenstates, $\tilde q'_{\alpha}$, are related by, $\, \tilde q'_{\alpha} = \sum  {R_{\alpha \beta}^{(q)}} \tilde q_{\beta}$, where,
$\tilde u'_{\alpha} 
= \left( \tilde c_L \, , \, \tilde c_R \, , \, \tilde t_L \, , \, \tilde t_R  \right)
$ , 
$\tilde d'_{\alpha} 
= \left(\tilde s_L \, , \, \tilde s_R \, , \, \tilde b_L \, , \, \tilde b_R \right)$
,
$ \tilde u_{\alpha} 
= \left( \tilde u_1 \, , \, \tilde u_2  \, , \,\tilde u_3  \, , \,\tilde u_4  \right) $
and
$\tilde d_{\alpha} 
= \left( \tilde d_1 \, , \, \tilde d_2 \, , \, \tilde d_3 \, , \, \tilde d_4  \ \right)$
.
%
%
%


We present next our results for the partial decay widths, containing the one-loop corrections in terms of the form factors $F_{L,R}^{qq'} (H_a)$ associated to each decay $H_a \to q \bar q'$, with $H_a=h_o, H_o, A_o$, and defined by $i F = -ig \bar u_q (p_1) (  {F_L^{qq'} (H)} P_L +  {F_R^{qq'} (H)} P_R) v_{q'}(p_2) H(p_3) $.
%
%
%
Similarly to the flavour preserving case, there are two types of one-loop diagrams that contribute, the triangular vertex loop diagrams and the FC self-energies of the external quarks. 
We show here only $F_{L}^{bs} (H_a)$ as an illustrative example (the rest can be found in~\cite{Curiel2}), \\

${ F_L^{bs} (H_a)}  = - \frac{{g_{H_a \tilde d_{\alpha} \tilde d_{\beta}}}}{ig} \frac{2 \alpha_s}{3 \pi} 
 \left(m_b  {R_{3 \alpha}^{(d)} R_{1 \beta}^{(d)*}}  (C_{11} - C_{12}) 
 + m_s  {R_{4 \alpha}^{(d)} R_ {2 \beta}^{(d)*}}  C_{12} + \right. $ \\
$\left. + {M_{ \tilde g}} {R_{4 \alpha}^{(d)} R_{1 \beta}^{(d)*}} C_0 \right) (m_b^2,m_{H_a}^2,m_s^2, {M_{\tilde g}^2}, {M_{\tilde
 d_{\alpha}}^2}, {M_{\tilde d_{\beta}}^2})
  + \frac{G_{H s \bar s}}{i g} \frac{m_b}{m_b^2- m_s^2} \kappa_L^d \left[ m_b (\Sigma_R^{bs} (m_b^2) + \right.$\\
 $ \left. + \Sigma_{Rs}^{bs} (m_b^2)) +  
 + m_s (\Sigma_L^{bs} (m_b^2) + \Sigma_{Ls}^{bs} (m_b^2)) \right] 
+ \frac{G_{H b \bar b}}{i g} \frac{1}{m_s^2 - m_b^2} \kappa_L^d \left[ m_s^2 \Sigma_L^{bs} (m_s^2) + \right.$\\
$ \left. + m_b m_s (\Sigma_{Rs}^{bs} (m_s^2) + \Sigma_R^{bs} (m_s^2))
 + m_b^2 \Sigma_{Ls}^{bs} (m_s^2) \right]$

\begin{center}
\begin{figure}
  \includegraphics[height=.15\textheight]{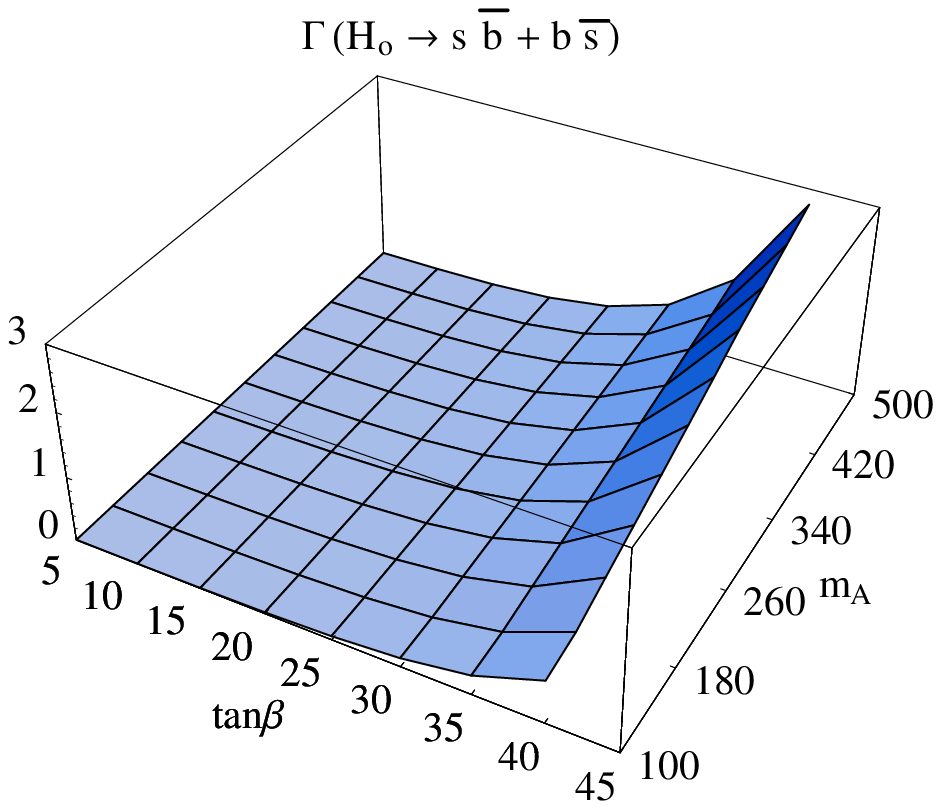}
  \includegraphics[height=.15\textheight]{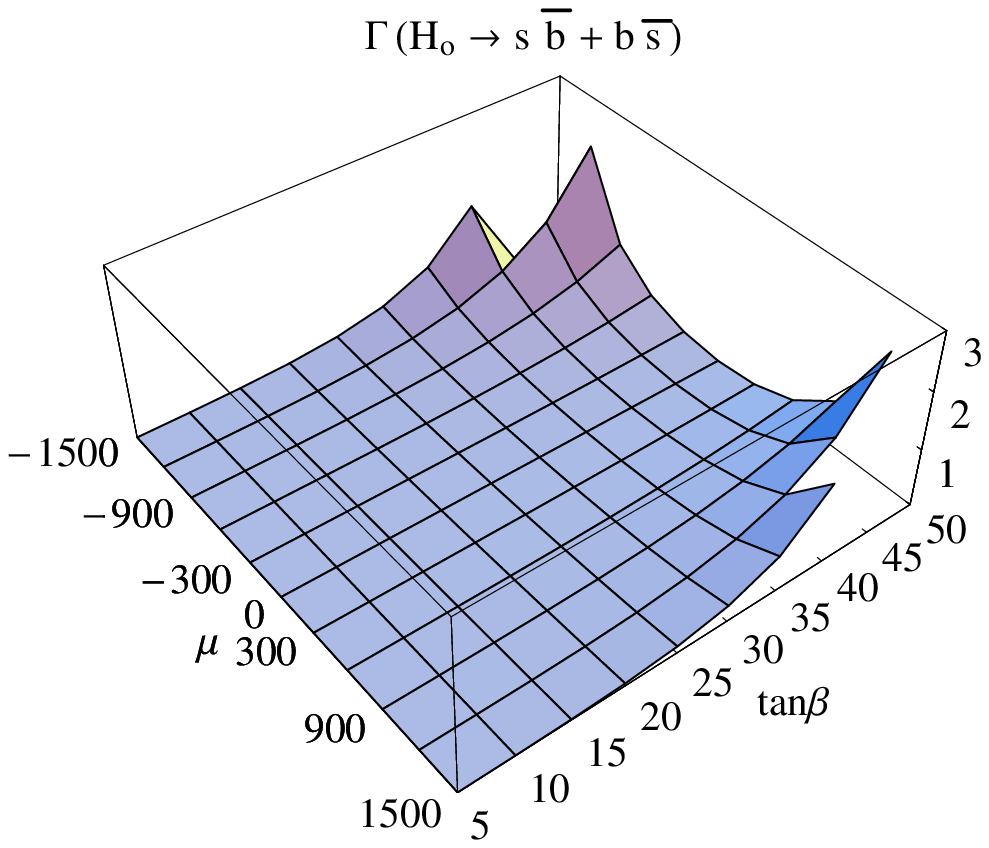}
  \includegraphics[height=.15\textheight]{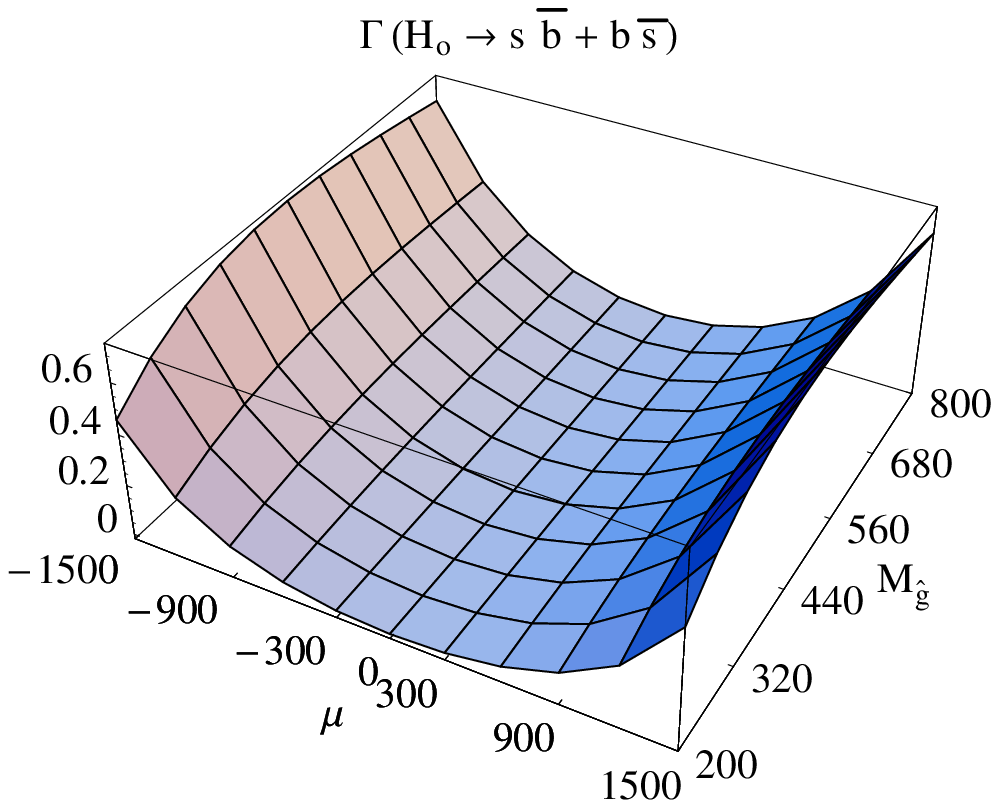}
  \includegraphics[height=.15\textheight]{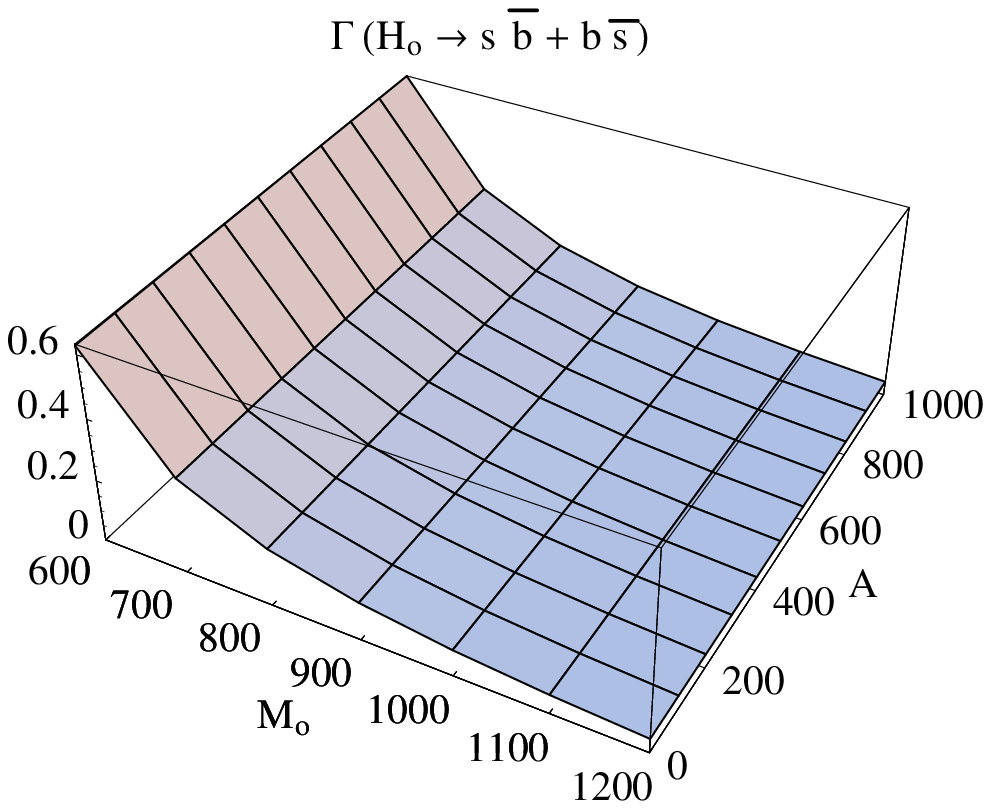}
\caption{$\Gamma (H_o \to b \bar s + s \bar b)$ in  $GeV$ as a function of the MSSM parameters. The corresponding fixed values are $\mu=1500 GeV$, $M_o=600 GeV$,
$M_{\tilde g}= 300 GeV$, $A = 200 GeV$, $m_A = 250 GeV$, $\tan \beta = 35$,
$ \lambda = 0.5$.}
\label{contour1}
\end{figure}
\end{center}
\begin{center}
\begin{figure}
  \includegraphics[height=.2\textheight]{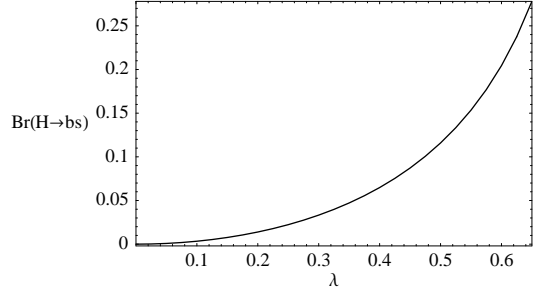}
 \caption{$Br(H_o \to b \bar s + s \bar b)$ as a function of $\lambda$ for the selected MSSM parameters of fig.~\ref{contour1}.} 
\label{branching1}
\end{figure}
\end{center}

Notice that the results for the form factors are finite, as expected, since renormalization is not needed in this process. The dependences on the MSSM parameters, $m_A,\tan \beta,\mu,M_{\tilde g},M_0,A$, are similar to the flavour preserving cases studied before. Here we assume universal squark masses $M_o$ and trilinear couplings, A. The dependency on the crucial flavour mixing parameter $\lambda$ enters in the rotation matrices and in the physical squark masses. 
%
We show in figs.~\ref{contour1} and~\ref{branching1} the numerical results for the case $H_o \to s \bar b + b\bar s$ as a function of the six previous MSSM parameters and $\lambda$. We see clearly that the branching ratio can be quite sizable. Similar results are found for $A_o$ and $h_o$ and also for the decays into $t \bar c$~\cite{Curiel2}. 
%
%
%

\vspace{-1.5cm}
%
Regarding the large $M_{SUSY}$ behaviour of these corrections, we find the following result for the leading ${\mathcal{O}}(\frac{M_{EW}}{M_{SUSY}})^0$ term of the expansion, valid for all $\tan \beta$ values,\\
$F_{L,R}^{bs} (\{h_o,H_o.A_o\}) = \frac{\alpha_s}{6 \pi} \frac{m_{b,s}}{2 m_W} \{\frac{\sin \alpha}{\cos \beta}, \frac{- \cos \alpha}{\cos \beta}, \mp i \tan \beta\} (\tan\beta + \{\cot\alpha, -\tan \alpha, \cot \beta\}) \frac{{\mu} {M_{\tilde g}}}{{M_o^2}} F(\lambda)$
%
%
where, $F(\lambda) = \frac{2}{\lambda^2} [(\lambda +1)\ln(\lambda+1) + (\lambda -1)\ln(1-\lambda) - 2\lambda]$. Similar formulas are found for the u-sector~\cite{Curiel2}.
%
%
Again, there is non-decoupling with $M_{SUSY}$ and the effect is enhanced at large $\tan \beta$. Decoupling appears only in the case of the light 
Higgs if $m_A\gg m_Z$, since $\cot\alpha \rightarrow -\tan\beta$, and thus $F_{L,R}^{bs} (h_o) \rightarrow 0$, recovering the SM result.

This SUSY non-decoupling behaviour explains the large FCHD rates found: 
$ BR(H_0,A_0\rightarrow b\bar s+s\bar b) \le 0.2$, 
$ BR(h_0\rightarrow b\bar s+s\bar b) \le 0.01$,
$ BR(H_0,A_0\rightarrow t\bar c+c\bar t) \le 5 \times 10^{-5}$,
 for $\lambda\le 0.6$ that will certainly produce interesting phenomenological SUSY signals at colliders.

\section{Non-decoupling of SUSY loops in effective Higgs-quark-quark
interactions}
Here we present the results for the effective Yukawa couplings generated from squark-gluino loops~\cite{Dobado_Temes}.
We start with the relevant terms of the MSSM classical action, containing the free part and the interaction terms,  
$ {S} = S_0 [H] + S_0 [q] + S [H,q] + 
 {S_0 [\tilde q]} +  {S_0[\tilde g]} +  {S [H,\tilde q]} 
+  {S [\tilde q, \tilde g, q]}, $
, and compute the one-loop effective action, at order $\alpha_S$, that is generated
after integration of squarks and gluinos in the path integral formalism, 
$e^{i  {\Delta \Gamma_{eff}[H,q]}}=\int  {[d\tilde Q^\dagger][d\tilde Q ]
[d\tilde g]} e^{i(  {S_0[\tilde q]} +  {S_0[\tilde g]}
+  {S[\tilde g, \tilde q, q]+S[H,\tilde q]})}$. We then perform a large 
$M_{SUSY}$ expansion and get 
the corresponding effective lagrangian, defined by,
$ \Delta{\Gamma_{eff} [H,q]} \equiv \int dx  
\Delta{{\mathcal L}_{eff} (H,q)} $.
After redefining properly the quark wave functions and quark masses we get,\\
 $\Delta{\mathcal L}_{eff} (H,q) =  
 {\Delta \lambda^{h^o\bar bb}}  { h^o \bar b b} 
+   {\Delta \lambda^{h^o\bar tt}}  {h^o \bar t t} \nonumber 
 +  {\Delta \lambda^{H^o\bar bb}}  { H^o \bar b b} 
+  {\Delta \lambda^{H^o\bar tt}}  {H^o \bar t t} 
+  {\Delta \lambda^{A^o\bar bb}}  { A^o \bar b b} 
+  {\Delta \lambda^{A^o\bar tt}}  { A^o \bar t t } \nonumber 
+  {\Delta \lambda^{H^+\bar t_L b_R}}  {(H^+ \bar t_L b_R + h.c.
)}
+  {\Delta \lambda^{H^+ \bar t_R b_L}}  { (H^+ \bar  t_R b_L + h.c.
)}$
%
with,\\
${\Delta \lambda^{h^o\bar bb}} \equiv \frac{m_b \sin \alpha}{ \cos \beta} G \left[\tan\beta
+ \cot\alpha \right] \, , \,
 {\Delta \lambda^{h^o\bar tt}} \equiv - \frac{ m_t \cos \alpha}{ \sin \beta} G \left[\cot\beta
+ \tan\alpha \right]  \\
 {\Delta \lambda^{H^o\bar bb}} \equiv - \frac{m_b \cos \alpha}{ \cos\beta} G \left[ \tan\beta
- \tan\alpha \right] \, , \,
 {\Delta \lambda^{H^o\bar tt}} \equiv - \frac{ m_t \sin \alpha}{ \sin\beta} G \left[ \cot\beta
- \cot\alpha \right] \\
 {\Delta \lambda^{A^o\bar bb}} \equiv 
 {\Delta \lambda^{A^o\bar tt}} \equiv 
 i  m_t \cot\beta  G \left[ \tan\beta
+ \cot\beta \right] \\
 {\Delta \lambda^{H^+\bar t_L b_R}} \equiv 
 {\Delta \lambda^{H^+\bar t_R b_L}} \equiv 
 m_t\sqrt{2} \cot\beta G  \left[ \tan\beta
+ \cot\beta \right]$ ; $G \equiv - \frac{\alpha_s}{3 \pi} \frac{g}{2 m_W} \frac{M_{\tilde g} \mu}{M_{SUSY}^2}$.
%

The important result is that new  $Hq\bar q$ couplings emerge 
of 2HDMIII type. These are precisely the   
non-decoupling effects from the squark-gluino loops. On the other hand, 
these exlicit expressions 
of the effective Yukawa couplings, which are valid for all 
$\tan\beta$ values, have interesting applications for
phenomenology as we will show next. 

\section{How to look for indirect SUSY signals}

Finally, we shortly review the main phenomenological consequences for future colliders. In order to do that, we have chosen a set of observables,

$\frac{BR(h^o \rightarrow b\bar b)}{BR(h^o \rightarrow \tau^+ \tau^-)}
\,\,\, , \,\,\,  
 \frac{BR(H^o \rightarrow b\bar b)}{BR(H^o \rightarrow \tau^+ \tau^-)}$
 $\frac{BR(A^o \rightarrow b\bar b)}{BR(A^o \rightarrow \tau^+ \tau^-)}
 \,\,\, , \,\,\,  
 \frac{BR(H^+ \rightarrow t\bar b)}{BR(H^+ \rightarrow \tau^+ \nu)}$
 $\frac{BR(t \rightarrow H^+ b)}{BR(t \rightarrow W^+ b)}$
 that are optimal for various reasons~\cite{Curiel1}. First, the dominant SUSY-QCD corrections do not decouple in the numerator. The SUSY-EW corrections do not decouple either (in the numerator and the denominator), 
 but are smaller~\cite{Carena,Guasch}. For example, for $\tan \beta=30$ and universal large $M_{SUSY}$, $\Delta_{SEW}\simeq 8\%,\,
 \Delta_{SQCD}\simeq -40\%$. There are sizeable SUSY-QCD corrections at large $\tan \beta$ in these observables and the ratios allow to cancel the production uncertainties and to minimize the systematic errors using the leptonic decays as control channels. 
Another interesting feature of these observables is that they will be experimentally accessible at LHC/TeVatron/Linear Colliders, being the SUSY corrections larger than the expected precision. 

\begin{figure}
  \includegraphics[height=.3\textheight]{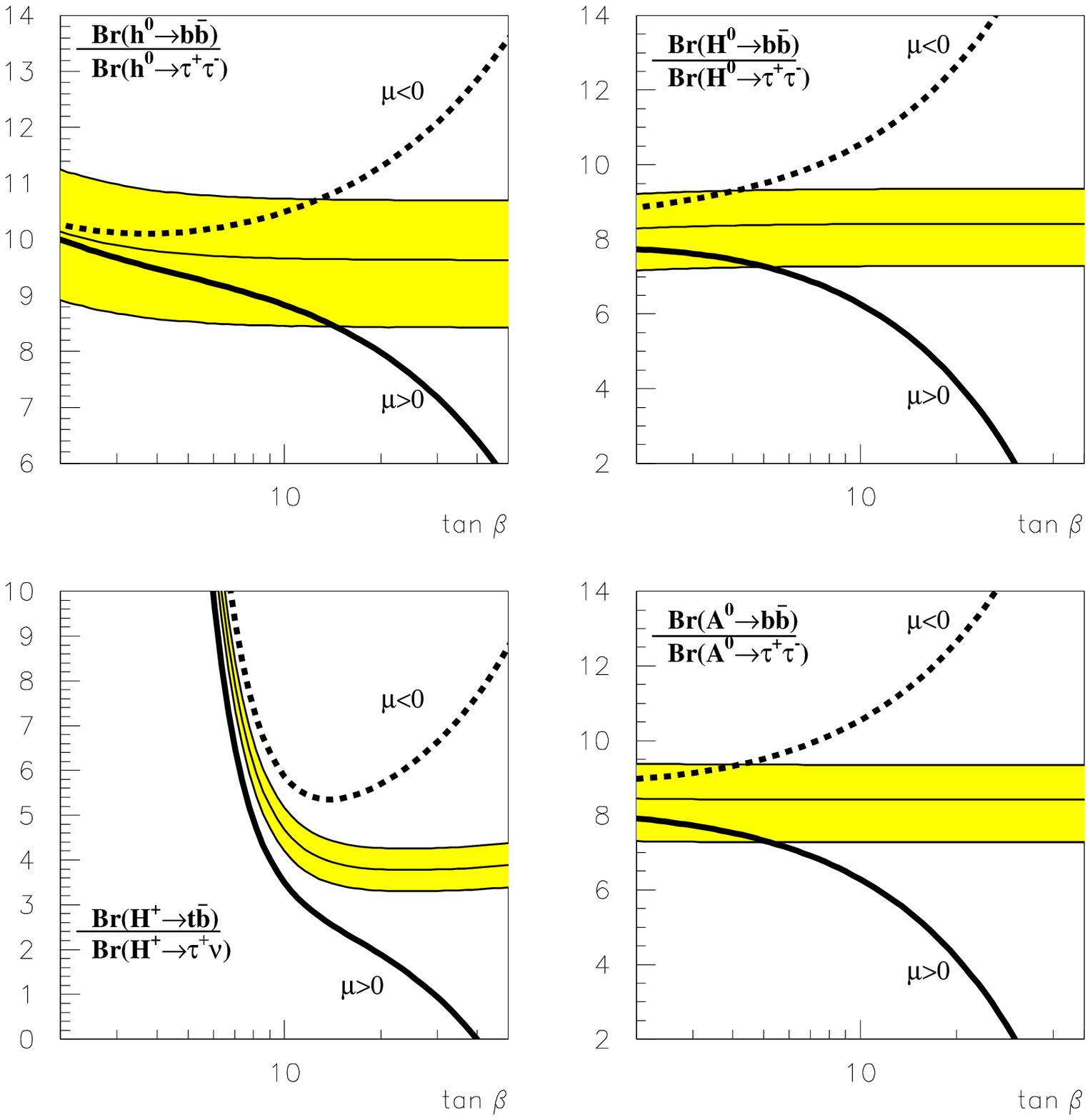}
  \includegraphics[height=.3\textheight]{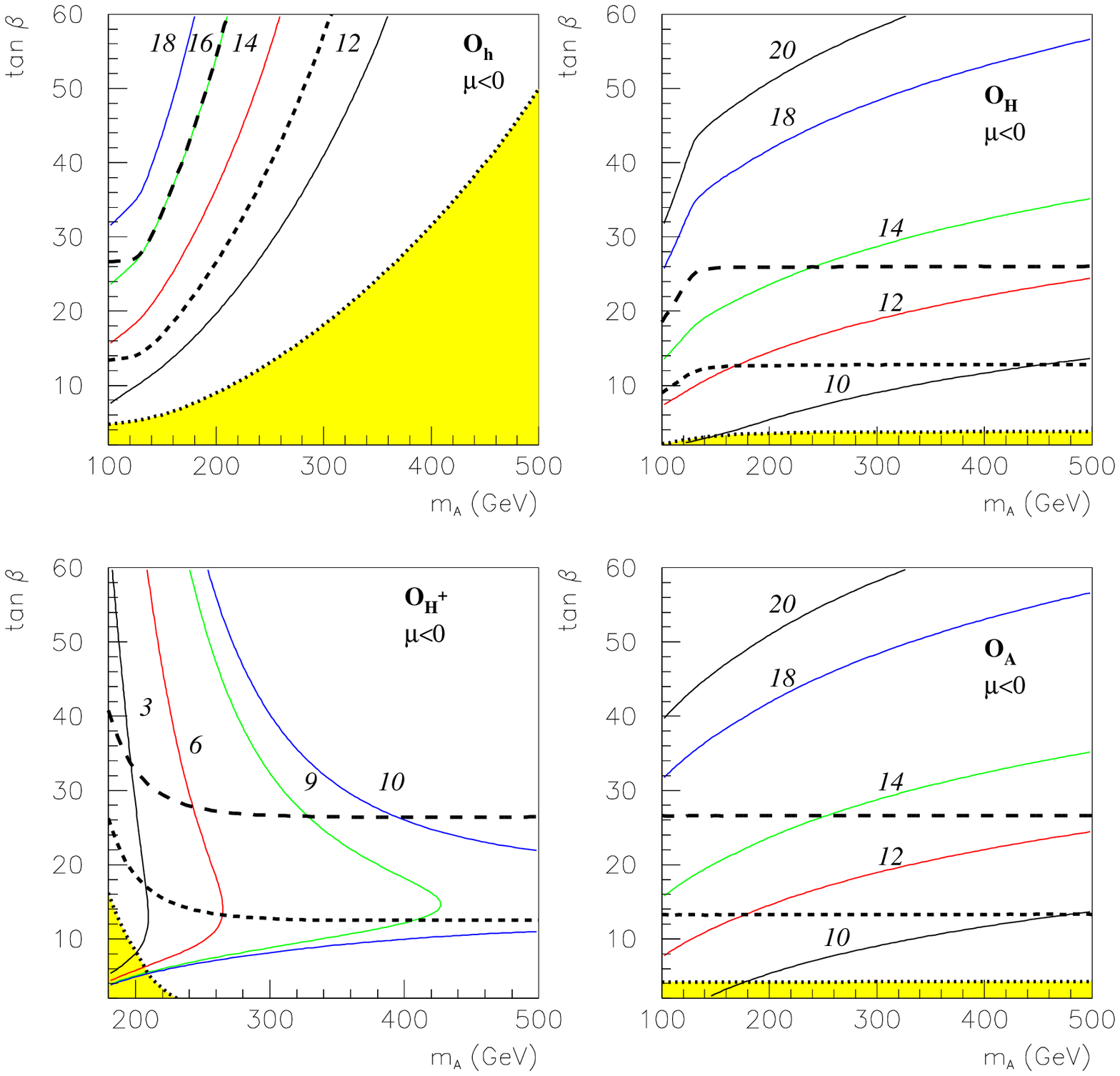}
\caption{(a){\it(Four left pannels)} Sensitivity to SUSY-QCD corrections vs. $\tan \beta$ for $m_A = 250$ GeV, $m_b= 4.6 \pm 0.2 \,GeV,
m_t= 174.3 \pm 5.1 \,GeV,\alpha_S (M_Z)= 0.118  \pm 0.002 $; (b){\it(Four right pannels)} Predictions for the observables including the SUSY-QCD corrections, as a function of $m_A$ and $\tan \beta$,
for $M_{\rm SUSY}=M_{\tilde g}=|\mu|$ and $\mu<0$. (Similar results for $\mu >0$).}
\label{finalhiggs}
\end{figure}

We show the sensitivity to SUSY-QCD corrections with $\tan \beta$ in fig.~\ref{finalhiggs}(a). The central line corresponds to the value of the observable without SUSY-QCD corrections, while the other two lines are our predictions for the observables, including the SUSY-QCD correction. The central band corresponds to the theoretical uncertainty in the observable coming from the experimental uncertainty in the SM parameters, which is
dominated by $\Delta m_b$. We can see that we are sensitive to the SUSY-QCD corrections for most of the $\tan \beta$ parameter space, having higher sensitivity at large $\tan \beta$~\cite{Curiel1,Guasch}.
 In fig.~\ref{finalhiggs}(b) we plot the predictions for the observables including the SUSY-QCD corrections as a function of $m_{A^o}$ and $\tan\beta$. The solid contour lines follow the points in the $(m_{A^o},\tan\beta)$ plane with constant
value of the corresponding observable, denoted here by $O$. The shaded area represents the region where the corrections are smaller than the mentioned theoretical uncertainty. The long (short) dashed lines join the points where an experimental resolution of $50\% (20\%)$ is required to achieve a meaningful measurement.The regions above these dashed lines fulfil the required sensitivity to the SUSY-QCD corrections. We see that the SUSY effects are visible in most of the
 parameter space even for the pessimistic case where only a 50\% resolution would be achieved. As already said before, there is greater sensitivity at large $\tan\beta$. For the light Higgs, to distinguish the MSSM and SM will only be possible at low  $m_A$.


\section{Conclusions}
In summary, we find a non-decoupling
behaviour of squark-gluino loops in $h^o \to \bar bb$ and $H^+ \to t\bar b$
and also in FCHD, $h^o,H^o,A^o \to \bar b s,s \bar b$ and $H^o,A^o \to t \bar c, c \bar t$. We compute the low energy effective theory
by integrating out squarks and gluinos in the path integral and find a  2HDM of type III, giving the explicit expressions 
to ${\mathcal O}
(\alpha_S)$ for the effective  $Hqq$ Yukawa couplings. Finally, we study the sensitivity to these non-decoupling effects at colliders with a set of optimal observables, concluding that they are promising to look for indirect SUSY signals, even if the SUSY spectrum is very heavy. 



\begin{theacknowledgments}
It is a pleasure to congratulate Augusto Garc\'\i a and Arnulfo Zepeda for their 60th birthday, and M.~J.~Herrero wishes to thank the organizers, specially to Miguel Angel P\'erez for inviting her to a very enjoyable meeting. She also thanks to her collegues A.~Dobado, H.~Haber, W.~Hollik, H.~Logan, S.~Pe\~naranda, S.~Rigolin and J.~F.~de Troc\'oniz for the various fruitful collaborations, which have been the basis for this summary talk.
\end{theacknowledgments}

\end{document}